\documentclass[fleqn,usenatbib]{mnras}

\usepackage{newtxtext,newtxmath}

\usepackage[T1]{fontenc}

\DeclareRobustCommand{\VAN}[3]{#2}
\let\VANthebibliography\thebibliography
\def\thebibliography{\DeclareRobustCommand{\VAN}[3]{##3}\VANthebibliography}

\usepackage{graphicx}	
\usepackage{amsmath}	
\usepackage{pdflscape}

\title[NGC5846\_UDG1]{Revisiting the distance and the globular cluster system of the remarkable galaxy UDG1 in the NGC 5846 group}

\author[Forbes et al.]{
Duncan A. Forbes,$^{1}$\thanks{E-mail:duncan.forbes@gmail.com (DAF)}
Bas van Heumen$^{1}$
and
Yimeng Tang$^{2}$
\\
$^{1}$ Centre for Astrophysics \& Supercomputing, Swinburne University, Hawthorn, VIC 3122, Australia\\
$^{2}$ Department of Astronomy \& Astrophysics, University of California Santa Cruz, 1156 High Street, Santa Cruz, CA 95064, USA
}

\date{Accepted XXX. Received YYY; in original form ZZZ}

\pubyear{\the\year{}}

\begin{document}
\label{firstpage}
\pagerange{\pageref{firstpage}--\pageref{lastpage}}
\maketitle

\begin{abstract}

Two studies that utilised the same HST/WFC3 imaging of NGC5846\_UDG1 have reported quite different total counts for its globular cluster (GC)  system, i.e. 54 $\pm$ 9 vs 33 $\pm$ 3 GCs. In both cases they counted all GCs, that met their selection criteria, down to the faintest magnitudes. 
They also disagree as to whether NGC5846\_UDG1 lies in the NGC 5846 group or well outside the group, in the field. As an ultra diffuse galaxy with one of the richest GC systems known, and therefore implications for its halo mass, it is important to understand which of these is closer to the truth. 
Here we present a new SBF-based distance to NGC5846\_UDG1 from HST/ACS imaging of 26.5 $\pm$ 2.7 Mpc, which places it squarely within the NGC 5846 group. Using this distance we adopt the standard approach of only counting GCs brighter than the turnover magnitude. This has the advantage of considering only the brighter GCs which are resolved in HST imaging and largely confirmed by spectroscopy, while also avoiding the fainter candidates for which contamination is potentially an issue. With this robust approach we find that the two studies are entirely consistent with each other. Both imply a total GC system of around 50 GCs and by inference a massive galaxy halo of greater than 10$^{11}$ M$_{\odot}$.  
We also revisit the two previous photometric studies focusing on half a dozen intermediate magnitude objects that are selected by one study but excluded by the other. These objects have GC-like magnitudes, sizes and are nearly round with GC-like appearances. They are very unlikely to be background galaxies or interloper GCs and thus bona fide GCs associated with NGC5846\_UDG1.

\end{abstract}

\begin{keywords}
galaxies: dwarf -- galaxies: haloes  -- galaxies: star clusters: general. 
\end{keywords}



\section{Introduction}

Globular clusters (GCs) can reveal much about their host galaxy, e.g. \cite{2006ARA&A..44..193B} and 
\cite{Beasley2020}.
For example, counting the total number of GCs associated with a galaxy can provide an estimate of its total halo mass (e.g. 
\cite{1997ApJ...481L..59B}; 
\cite{2009MNRAS.392L...1S}; 
\cite{2015ApJ...806...36H}; \cite{Forbes2018}; 
\cite{Burkert2020};  \cite{2025ApJ...988...70D},   
which can otherwise be very difficult to measure for an individual galaxy. Ultra Diffuse Galaxies (UDGs) are particularly interesting in this context. They appear to also follow the GC halo mass relation of normal galaxies  
\citep{Forbes2025b},  
despite hosting more GCs, on average, than classical dwarf galaxies of the same stellar mass. Equivalently, UDGs can have on average higher GC system mass to galaxy stellar mass ratios (M$_{GC}$/M$_\ast$). For example, in the Coma cluster classical dwarfs with stellar masses of $\sim$10$^8$ M$_{\odot}$ have up 10-15 GCs and M$_{GC}$/M$_\ast$ up to 2-3\%. Whereas similar mass Coma UDGs have up to 50-80 GCs and ratios up to $\sim$10\% \citep{Forbes2020}. The Perseus cluster also hosts 
many GC-rich UDGs as found by \cite{Janssens2024} and \cite{Saifollahi2025}. 
See also works by \cite{2025ApJ...988...70D}. 
However, in low density environments UDGs and classical dwarfs tend to be similarly GC-poor with low 
M$_{GC}$/M$_\ast$ ratios 
(see review by \citealt{2026arXiv260221875G}). 
In the MATLAS survey, the majority of their 74 field and group UDGs imaged by HST revealed fewer than two GCs, with only three galaxies having more than 20 GCs 
\citep{Marleau2024}. 

Here we focus on a remarkable UDG with a very rich GC system and one of the highest M$_{GC}$/M$_{\ast}$ ratios known. In 2022, \cite{Danieli2022} noted that  
{\it No other known galaxy has such a striking,
dominant globular cluster population.}
It is especially notable because it lies outside of a cluster. 
The galaxy was originally identified as NGC5846-156 by \cite{2005AJ....130.1502M}. Later,  it was determined to be an UDG with a populous GC system by \cite{Forbes2019}
who named it NGC5846\_UDG1. It was also identified in the MATLAS survey (named MATLAS-2019) and found to be the UDG with the most populous GC system in that survey of low density environments \citep{Marleau2024}. 
This galaxy (which we refer to here as UDG1) has effectively become the `posterchild' for a GC-rich UDG with several imaging studies of its GC system and two conducting spectroscopic followup of its GCs.
We refer the reader to figure 1 of \cite{Danieli2022} which visually highlights the plethora of compact sources around UDG1 (and their relative absence in the rest of the HST image).


However, despite this attention in the literature, there is some disagreement as to the richness of its GC system (and hence its total halo mass). A brief history of UDG1: 
The first study used deep multi-filter imaging from the VST \citep{Forbes2019}. From these  ground-based data, \cite{Forbes2021}, assuming a distance of 24.9 Mpc, 
estimated a total GC system of 45 (with no uncertainty quoted given the limitations under ground-based  seeing). At such distances, imaging with HST can resolve GCs. So in addition to magnitude and colour, subsequent HST studies of UDG1's GC system could include morphological properties such as GC size, ellipticity and concentration. Thus foreground stars can be effectively identified and removed from any GC candidate list. 

Using HST/ACS imaging, \cite{Muller2021} derived a total GC count of 26 $\pm$ 6 and 36 $\pm$ 6 depending on whether a Bayesian approach was adopted or not. Assuming a GC luminosity function (GCLF) turnover of M$_V$ = --7.6 \citep{2012Ap&SS.341..195R}, they fit the observed GCLF and found a distance of 20.7$^{+2.3}_{-2.1}$ Mpc. Using deeper and higher resolution HST/WFC3 imaging, \cite{Danieli2022} derived a total GC count of 54 $\pm$ 9. Their resulting GCLF was fully consistent with a GCLF turnover of M$_V$ = --7.5 for an assumed distance of 26.5 Mpc.
Also using HST/ACS imaging, \cite{Marleau2024} derived a GC count of 38 $\pm$ 7 for an assumed  distance of 20.3 Mpc (i.e. well outside of the NGC 5846 group) and a GCLF of M$_V$ = --7.3. Most recently, \cite{2025arXiv251209990G} combined new deep u band imaging from the GTC, under 1 arcsec seeing, with the existing HST data. They derived a total count of 33 $\pm$ 3 GCs and noted that some properties of their selected GCs favoured a 
distance of 20.0 $\pm$ 0.9 Mpc. 

The history of estimating the total GC counts for UDG1 in the literature is summarised in Table 1. It is clear that the uncertainties quoted by the various studies are measurement errors and do not appear to include systematic effects in terms of selection criteria and/or the distance uncertainty.  

During these imaging campaigns, spectra were obtained of several GC candidates by \cite{Muller2020} using VLT/MUSE and by \cite{2025MNRAS.539..674H} using Keck/KCWI. From these works there are now 20 GCs confirmed with radial velocities. This is the largest set of confirmed GCs for any UDG. We note that in both studies, none of the GC candidates from HST imaging were found to be background galaxies.

Surprisingly, the two studies using the deepest and highest resolution data have reached contradictory conclusions. One suggested that UDG1 has a remarkable 54 GCs (around 1/3 of the Milky Way's GC count, with around 1/300 of its stellar mass) and is located within the NGC 5846 group \citep{Danieli2022}. Alternatively, and perhaps even more remarkably, it has been suggested UDG1 could host 33 GCs and located outside of the group, in the field \citep{2025arXiv251209990G}. 
Given the key nature of UDG1, its focus in the literature and 
its remarkable GC system,  
it is important to `solve the mystery' and determine which of these situations is most likely. 

Here we start by revisiting the GC system of UDG1 focusing on the 
\cite{Danieli2022} and \cite{2025arXiv251209990G} studies.
Both studies used HST imaging 
which was deep enough to detect essentially all GCs at the distance of UDG1.
Thus the total number of GC candidates can essentially be tallied-up from the HST imaging without knowing the exact distance to UDG1. However, the two studies disagree on the selected GC candidates, leading to total GC counts that are quite discrepant (i.e. 54 vs 33 GCs). 
We attempt to understand the differences (and similarities) between the two studies to help guide future GC studies. 
We present colour-magnitude diagrams to illustrate differences in the selection process and focus on half dozen GCs included in one study but excluded by the other. 

However, our main result is based on a new surface brightness fluctuation (SBF) distance to UDG1. Using this distance and the conventional approach of only counting GCs brighter than the turnover, we derive a total number of GCs from each study. We find both studies to be consistent with a system of $\sim$50 GCs. 

\begin{table}
\begin{center}
\begin{tabular}{lccccc}
  \hline
 Study & N$_{GC}$ & Dist. (Mpc) & Res. (${''}$) & Exposure (s) & Depth (mag) \\
 \hline
GA26 & 33$\pm$3 & 20.0 & 0.04 & 2360 & 27\\
M24 & 38$\pm$7 & 20.3 & 0.05 & 1854 & 26.5\\
D22 & 54$\pm$9 & 26.5 & 0.04 & 2360 & 27 \\
M21 &  36$\pm$6 & 20.7 & 0.05 & 1030 & 26\\
F19 & $\sim$45 & 24.9 & $\sim$1 & 11250 & 25.5\\ 
\hline
\end{tabular}
\caption{Literature studies of the globular cluster system of NGC5846\_UDG1. Columns are: study (GA26  \protect\cite{2025arXiv251209990G}, 
M24 
\protect\cite{Marleau2024}, D22 
\protect\cite{Danieli2022}, M21 
\protect\cite{Muller2021}, F21 
\protect\cite{Forbes2021}), total number of globular cluster (for M21 the non-Bayesian approach is listed), the quoted distance, the effective resolution, the exposure time, and the approximate point source limit in the F606W filter (or g band for F21). 
}
\end{center}
\end{table}

\begin{figure*}
    \centering
    \includegraphics[width=0.99\linewidth]{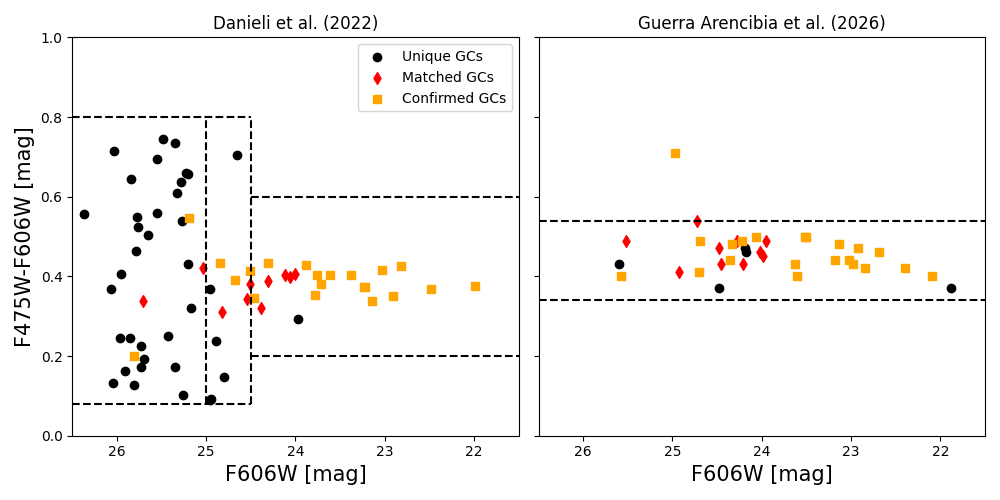}
    \caption{Colour-magnitude diagram for GCs in UDG1 as selected by 
    \protect\cite{2025arXiv251209990G} (right) and 
    \protect\cite{Danieli2022} (left). Their F606W magnitude and/or colour selection is indicated in each panel as dashed lines. GCs are colour coded by spectral confirmation, matched between studies or unique to one study.  All GCs brighter than F606W = 25 mag. are resolved by HST/WFC3. The photometry of 
    \protect\cite{2025arXiv251209990G} reveals a systematic trend with magnitude, whereas 
    \protect\cite{Danieli2022} find a near constant colour for the brighter GCs.  
}
    \label{fig:cmd}
\end{figure*}

\section{Similarities and Differences}

Both the \cite{Danieli2022} and \cite{2025arXiv251209990G} studies 
used the HST/WFC3 F475W and F606W (hereafter F475 and F606) data, which reach a depth of F606 $\sim$ 27 mag., i.e. at least 2 mag. below the expected GCLF turnover for UDG1. Whereas the latter study also incorporated 
HST/ACS F814W and ground-based OSIRIS u band imaging. 
Both employed magnitude, colour and size criteria to select GCs, with 
\cite{2025arXiv251209990G} also including ellipticity. They were both guided by the photometric properties of the spectroscopically-confirmed GCs. A key difference in approach is that \cite{2025arXiv251209990G} relied on their selection criteria to remove contaminants, whereas \cite{Danieli2022} used background fields within the same HST/WFC3 image. 

In order to calculate the total magnitude of a compact source in HST imaging, the standard approach is to measure an aperture magnitude and then make a correction to an infinite aperture following the documentation published by STScI. Such an approach was followed by \cite{Danieli2022}, as well as 
\cite{Muller2021} and \cite{Marleau2024}. \cite{2025arXiv251209990G} used GC radial profiles to obtain two aperture corrections, i.e from 2.7 pix (0.108 arcsec) to 0.6 arcsec radius ($\Delta$m $\sim$0.62 mag. for F606) and another to infinity ($\Delta$m = 0.06 mag.). We note there appears to be a typo in their table 3 which describes the first correction as ``0 to 0.60'' arcsec. Both studies applied the same Galactic extinction correction (e.g. A$_{F606}$ = 0.131). 

The 36 selected GC candidates of 
\cite{2025arXiv251209990G} with total magnitudes and other parameters 
are listed in their appendix C. Their list includes 4 candidates outside of 2~R$_e$ (the galaxy effective radius), with two of those showing  
star-like profiles. A  list of the \cite{Danieli2022} GC candidates and their measured properties within 2 $\times$ R$_e$ 
has been kindly supplied by S. Danieli (2025; priv. comm.). 

In the Appendix we compare the total F606 magnitudes from the two studies
relative to the F606 magnitudes from table 3 of \cite{Muller2021} who used the ACS F606 filter. Although we find some photometric differences this is not responsible for differences in reported 
GC numbers.

In principle, several aspects of the GC candidate selection process could lead to differences in the final GC count. For example, the radial extent of the GC system largely explains the lower GC counts of \cite{Saifollahi2022} compared to other literature works for UDGs in the Coma cluster (see 
\citealt{Forbes2024} for details). However, for UDG1  \cite{2025arXiv251209990G} and \cite{Danieli2022} are in good agreement finding the GC half number radius relative to the galaxy effective radius of R$_{GC}$/R$_e$ = 0.7 and 0.8 $\pm$ 0.1 (for R$_e$ = 16.0 $\pm$ 0.5 arcsec and 15.6 arcsec)  respectively.
This implies the vast bulk of the GC system lies within 2~R$_e$.

Other sources of potential difference are the corrections for magnitude and radial incompleteness. \cite{Danieli2022} found a background-corrected number of 50.4 GC candidates for  F606 $<$ 26.5 and within 2~R$_e$. They estimated that they missed 0.5 GC at fainter magnitudes and 3 GCs at larger radii. This gave them a total of 53.9 GCs for the UDG1 system. Similarly, \cite{2025arXiv251209990G} estimated 1 GC was missing due to magnitude incompleteness and they found 4 GC candidates beyond 2~R$_e$ (although two may be foreground stars). These small corrections can not explain the differences between the two studies. 

\cite{2025arXiv251209990G} additionally restricted GC candidates to have ellipticity between 0 and 0.2 (see their figure 2). Some 95\% of the Milky Way's GCs have ellipticities in this range, with a mean value of $\sim$0.07. However, for dwarf galaxies with weaker tidal fields one might expect more slightly diffuse and elongated GCs to be present. That indeed appears to be the case for GCs in the LMC and SMC with mean ellipticities of 0.16$\pm$0.08 and 0.19$\pm$0.06 respectively \citep{1996A&AS..116..447S}. We estimate that a less restrictive ellipticity criterion might have added one or two more GC candidates to their list. So again, this choice is not responsible for the large difference in the final GC counts.

So discrepancies in the final GC count does not appear to be due to differences in the photometric measurements or  depth, Galactic extinction applied, radial extent of the GC system, corrections for magnitude or radial incompleteness, or ellipticity. 
A significant difference between the two studies is however highlighted by \cite{2025arXiv251209990G}, i.e. their stricter colour selection of 
0.34 $<$ F475-F606 $<$ 0.54 (along with 0.26 $<$ F606-F814 $<$ 0.55 and  0.75 $<$ u-F475 $<$ 1.65).
They noted that if they relaxed their colour range to resemble that of \cite{Danieli2022}, they would have found a dozen more GCs, i.e. 43 $\pm$ 2 GCs within 2~R$_e$ compared to the 50.4 GCs found by \cite{Danieli2022}. In the next section we compare the colour-magnitude diagram (CMD) of GC candidates from both studies. 

\section{The Colour-Magnitude Diagram of the GC System}

In Fig. \ref{fig:cmd} we show the F475--F606 vs F606 CMD for both studies plotting their selected GC candidates and their selection criteria (as dashed lines) in this parameter space. The left hand panel is similar to fig. 2 of \cite{Danieli2022} except that we show their total rather than their AUTO magnitudes (the latter lacking aperture corrections). It only shows those GCs within 2~R$_e$ 
(an additional 3 GCs are expected at larger radii). As shown, their GC colour criterion is relaxed for F606 $>$ 24.5 mag. (but GC candidates are still required to be resolved). At fainter magnitudes, F606 $>$ 25 mag., there is no size criterion applied.

\cite{2025arXiv251209990G} did not show a CMD. Here, we show the CMD of their 36 GC candidates (i.e. including those that lie beyond 2~R$_e$ and with star-like profiles), in the right panel of 
Fig~\ref{fig:cmd}. Based on their photometric tests, their F475-F606 colour criterion is fixed for all magnitudes (they also employed other criteria not shown). 

\begin{figure}
    \centering
    \includegraphics[width=0.99\linewidth]{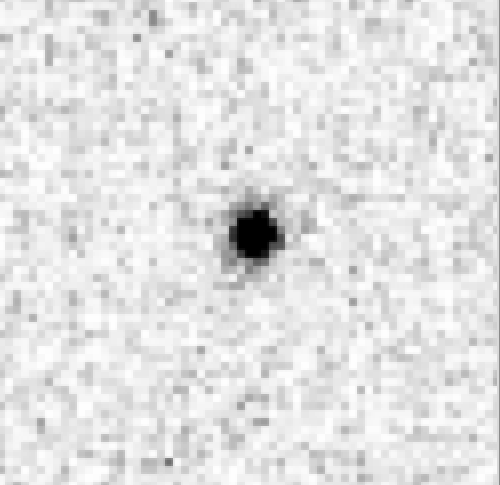}
    \caption{HST/WFC3 image in the F606 filter of a bright GC candidate. The image is approximately 2 $\times$ 2 arcsec with North up and East left. 
    This object was selected as a GC candidate by  
    \protect\cite{Danieli2022} measuring a magnitude of F606 = 23.97, colour of F475--F606 = 0.29 and FWHM size of 4.33 pix. This object was not selected by     \protect\cite{2025arXiv251209990G} as a GC candidate. The object is resolved and has a GC-like morphology. This object lies outside of the KWCI footprint of     \protect\cite{2025MNRAS.539..674H} and so lacks spectral confirmation.}
    \label{fig:gc1258}
\end{figure}

We now discuss the CMD of both studies in three magnitude ranges. We start with the brightest GCs (F606 $<$ 24.5 mag.) using the total magnitudes from each study (we remind the reader that the 
\cite{2025arXiv251209990G} magnitudes are systematically brighter by $\sim$0.1 mag., on average, than those of  \cite{Danieli2022}; see Appendix). 
In the left panel we show the GCs 
selected by \cite{Danieli2022} which are either confirmed by spectra or in common to both studies. They largely have near constant colour of F475--F606 $\sim$ 0.4. There is only one object unique to their study. In Fig.~\ref{fig:gc1258} we show the HST/WFC3 F606 image of this unique GC candidate (located at RA = 226.3267093, Dec = 1.8097772195). 
It is resolved (FWHM = 4.33 pix) and clearly has a GC-like morphology but it was not selected as a GC candidate by \cite{2025arXiv251209990G}, 
due to its blueish colour. 
The GC candidate lies outside of the KWCI footprint of \cite{2025MNRAS.539..674H} and so lacks spectral confirmation. However, it is likely a bona fide GC. 

In the right panel, the brightest GCs of \cite{2025arXiv251209990G} show a comparable colour but reveal a 
systematic trend to become redder by $\sim$0.1 mag. for fainter magnitudes.
Their 3 GC candidates unique to their study all lie outside of 2~R$_e$, with the brightest (GC35, F606 = 21.88 mag.) having a star-like profile. 
We remind the reader that the vast bulk of these bright objects are confirmed by spectra and that they represent the most luminous GCs, and hence the vast bulk of the GC system mass. Thus the ratio of total GC luminosity (mass) to galaxy luminosity (mass) from these GCs alone approaches $\sim$10\%, independent of the  GCs fainter than the turnover.
For comparison, if the luminosity of all 54 GCs, as determined by D22, are combined they give a V-band luminosity of 7.6 $\times$ 10$^6$ L$_{\odot}$. This represents a fraction of 12.9\% relative to the total galaxy luminosity, i.e. 5.9 
$\times$ 10$^7$ L$_{\odot}$. 


For faintest GC candidates (F606 $>$ 25 mag.) there is a large difference between the two studies. This is driven by the choice of using background fields 
\citep{Danieli2022} 
or relying on selection criteria \citep{2025arXiv251209990G}. 
\cite{2025arXiv251209990G} selected only 3 GC candidates at these faint magnitudes with 1 being  spectroscopically-confirmed, 1 in common with \cite{Danieli2022} and 1 unique GC candidate. No candidates with F606 $>$ 25.6 mag. were  selected. In contrast, \cite{Danieli2022} selected a large number of GC candidates, noting that around half may be  background contaminants (which they correct for in their final GC number). 
Given the statistical nature of the background contaminants at these faint magnitudes we don't discuss these GC candidates further other than to remind the reader that they are all projected within 2~R$_e$ of the centre of UDG1. 


It is more instructive to focus on the intermediate magnitudes (24.5 $<$ F606 $<$ 25). We note that the GC candidates in this range lie beyond the GCLF turnover, at fainter magnitudes, for all of the possible distances listed in Table 1.  In this magnitude range 
\cite{2025arXiv251209990G} have 5 GC candidates, 
with one of the spectroscopically-confirmed GCs lying outside of their colour criterion (which they suspect is due to an anomalous F475 magnitude). All 3 unique GCs lie beyond 2~R$_e$.

\begin{table}
\begin{center}
\begin{tabular}{lcccccc}
  \hline
 GC & RA & Dec & F606 & F475-F606 & FWHM\\
  & (J2000) & (J2000) & (mag) & (mag) & (pix)\\
 \hline
A & 226.34004205	& 1.8085222060	& 24.659	& 0.704 & 3.09\\
B & 226.34016774	& 1.8171241842	& 24.806	& 0.146 & 2.49\\
C & 226.33361727	& 1.8152357179	& 24.885	& 0.236 & 3.85\\
D & 226.33120177	& 1.8087172171	& 24.921	& 0.091 & 3.70\\
E & 226.33272246	& 1.8094940081	& 24.956	& 0.367 & 3.19\\
F & 226.32943095	& 1.8061618264	& 24.963	& 0.089 & 2.90\\
G & 226.33834602	& 1.8183536567	& 24.682	& 0.391	& 3.26\\
H & 226.33365171	& 1.8157238766	& 24.845	&	0.434 & 2.80\\
I & 226.33247812	& 1.8085526172	& 24.824	&	0.310 & 2.15\\
\hline
\end{tabular}
\caption{Resolved GC candidates in the range 24.5 $<$ F606 $<$ 25.0. GC candidates A-F were selected by 
\protect\cite{Danieli2022} but not by 
\protect\cite{2025arXiv251209990G}. GCs G and H are spectroscopically-confirmed. GC candidate `I' was selected by both studies. Coordinates, F606 magnitude and F475-F606 colour, FWHM size are supplied by S. Danieli (2025, priv. comm.). GCs G,H,I are matched with GC17, GC18, 
GC33 in 
\protect\cite{2025arXiv251209990G}. 
}
\end{center}
\end{table}

\begin{figure}
    \centering
    \includegraphics[width=0.99\linewidth]{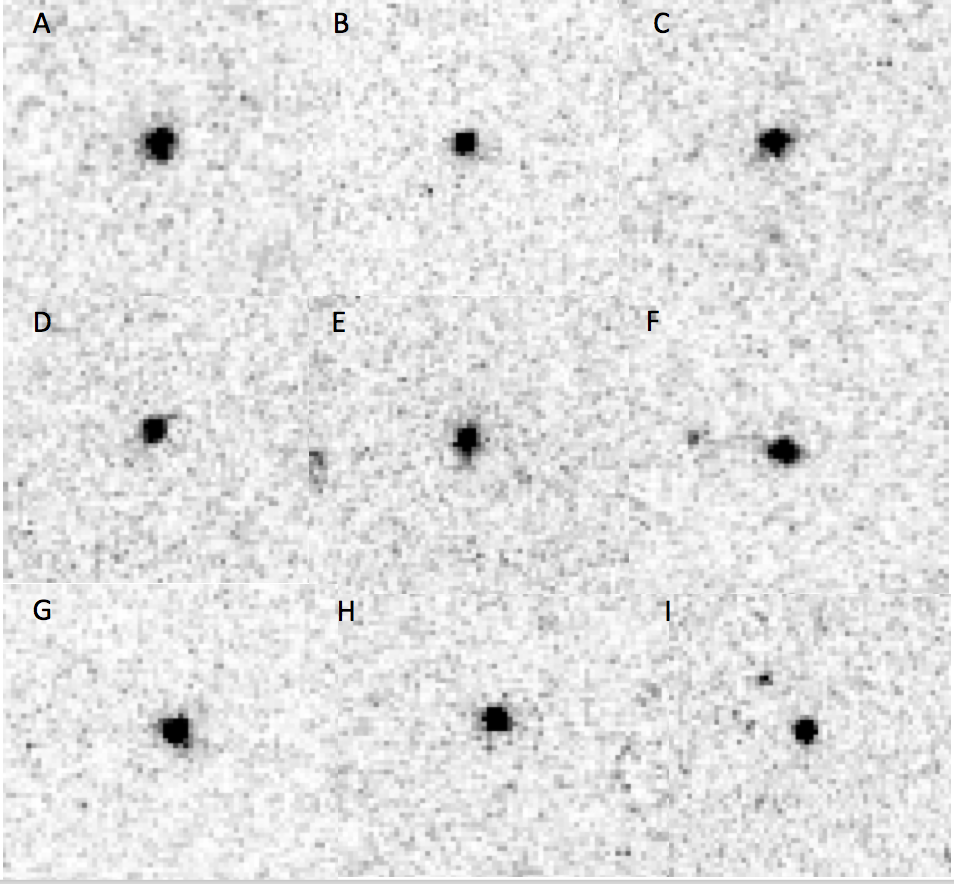}
    \caption{HST/WFC3 images in the F606 filter of GCs with 24.5 $<$ F606W $<$ 25.0. Individual images are approximately 2 $\times$ 2 arcsec with North up and East left.     
    The top two rows show the 6 GC candidates selected by 
    \protect\cite{Danieli2022} but not by 
    \protect\cite{2025arXiv251209990G}. The bottom row shows the two confirmed GCs and one additional GC candidate selected by both studies. All 9 objects are resolved by HST/WFC3 and have GC-like morphology. 
}
    \label{fig:mosaic}
\end{figure}

In this intermediate magnitude range \cite{Danieli2022} expands the colour selection range compared to the brighter GCs to 
0.08 $<$ F475--F606 $<$ 0.80 (to account for photometric errors) but still requires that they be resolved, i.e. sizes of 2.1 $<$ FWHM $<$ 4.5 pix. The latter ensures that they have sizes similar to the confirmed GCs (which have an average FWHM $\sim$ 3 pix). 
Their estimate of background contaminants in this region of the CMD is $\sim$ 0.7 objects. 
Their list 
includes two confirmed GCs, a common GC and 6 GCs unique to \cite{Danieli2022}. 
HST/WFC3 F606 images of these 6 GC candidates that were selected by \cite{Danieli2022} but not by \cite{2025arXiv251209990G} are shown in Fig.~\ref{fig:mosaic} denoted as A--F. 
Fig.~\ref{fig:mosaic} also shows two confirmed GCs (G and H) and one common GC (I) in this magnitude range (they are candidates GC17, GC18 and GC33 in \citealt{2025arXiv251209990G}). All 9 are notably GC-like in appearance. For each we list in Table 2 their coordinates, F606 magnitude, F475-F606 colour and FWHM size from the list supplied by S. Danieli (2025, priv. comm.).  

S. Guerra Arencibia (2026, priv. comm.) has identified these objects confirming they were rejected due to having colours outside of their colour criteria. 
Candidate F is also rejected being outside of their ellipticity criterion. Profile fitting indicates that candidate B has a star-like profile, while the others have profiles similar to the confirmed GCs, albeit at the more extended end of the distribution. 

If these GC candidates are not GCs associated with UDG1 as claimed by 
\cite{2025arXiv251209990G} the obvious question is what are they?
One object (B) has a star-like profile and a small FWHM size of 2.49 pixels (see Table 2). It is possible this is a foreground star.
However, the other GC candidates are well resolved in the WFC3 imaging so foreground stars can be ruled out. They could, in principle, be background galaxies. This could explain why their colours are different to those of the confirmed GCs (although generally redder colours would be expected of distant galaxies). 
Such galaxies would also have to be nearly-round with angular sizes that coincidentally have  GC-like sizes, of 3-5 pc, and GC-like magnitudes at the distance to UDG1.  They must also be projected within 2~R$_e$ of UDG1's centre. 
They are not all clustered together on the sky as might be expected for a background group of galaxies. We note that 
\cite{Danieli2022} estimated the contamination rate, based on background fields, in the 
24.5 $<$ F606 $<$ 25 magnitude range to be $<$1 background galaxy. 
We suggest that it is highly unlikely that they are nearly round background galaxies.


Could they be interloper GCs associated with the NGC 5846 group and not UDG1 itself? The frequency of interloper GCs from the large galaxies in the group, i.e. NGC 5846 itself, NGC 5813 and NGC 5838 at the location and velocity of UDG1 was estimated by 
\cite{2025MNRAS.539..674H} to be fewer than one GC. Thus `contamination' by half a dozen such GCs, within 2~R$_e$, is very unlikely. 

We conclude that the half dozen, intermediate magnitude GC candidates unique to \cite{Danieli2022} are indeed bona fide GCs of UDG1 and should not have been rejected by \cite{2025arXiv251209990G} despite having odd colours. We suspect that the spread in colours of these fainter GC candidates is due to increased photometric uncertainty (although ultimately spectra are required to confirm their true nature). 
We note that the faintest confirmed GC (with F606 $>$ 25 mag.) has a colour of F475-F606 = 0.4 as measured by \cite{2025arXiv251209990G}  and of 0.2 as measured by \cite{Danieli2022}. This suggests that 
the photometric errors may be larger than appreciated. 

Without a complete list of both GC candidates and rejected sources from both studies it is problematic to further analyse the differences in their methods. We now switch to an alternative approach to estimating the total GC count of UDG1 based on its brightest GCs and its distance.


\section{A New Distance to UDG1}

We now turn to the distance to UDG1, which has been the subject of debate in the literature. As summarised in Table 1, various distances to UDG1 (and hence different apparent magnitudes for the GCLF turnover) have been quoted. For the two studies re-examined here, an exact distance was not required in order to estimate the total number of GCs as the depth of the imaging was sufficient to detect all but one or two GCs. \cite{Danieli2022} showed that a NGC 5846 group distance of 26.5 Mpc \citep{2017ApJ...843...16K} was consistent with their GCLF. Whereas after creating their GC candidate list, \cite{2025arXiv251209990G} fit their GCLF and mean GC size to favour a distance of 20.0 $\pm$ 0.9 Mpc. 
Their closer distance would place UDG1 well outside of the NGC 5846 group, which has a virial radius of R$_V$ $\sim$ 0.6 Mpc, in the field. 

\begin{figure}
    \centering
    \includegraphics[width=1.1\linewidth]{figures/phase_space.png}
    \caption{Phase space diagram for galaxies in the NGC 5846 group from 
    \protect\cite{2010A&A...511A..12E} showing radial velocity vs projected radius. The location of UDG1 is also shown (red square). The histogram shows the distribution of radial velocities with UDG1 marked as a red line. The group velocity dispersion is 320 km/s. UDG1 is likely a NGC 5846 group member based on its location in the group phase diagram. 
}
    \label{fig:phase}
\end{figure}

We show in Fig. \ref{fig:phase} the phase space diagram for the NGC 5846 group. The data are taken from table 1 of \cite{2010A&A...511A..12E}. The plot shows galaxy radial velocity vs projected radius from NGC 5846,  including the location of UDG1. As noted by \cite{2005AJ....130.1502M} the surface density of galaxies drops off dramatically at a radius of 110 arcmins ($\sim$ 0.8 Mpc), which they associate with the second turnaround radius. They derive a group velocity dispersion of 320 km/s and a virial mass of 
8 $\times$ 10$^{13}$ M$_{\odot}$. UDG1 sits well within the second turnaround radius (and virial radius of 0.6 Mpc) and within 1.5$\times$ the velocity dispersion of the mean group velocity. This suggests a distance to UDG1 that is compatible with the group distance of 26.5 $\pm$ 0.8 Mpc \citep{2017ApJ...843...16K}.  

Perhaps the gold standard for determining distance to a galaxy like UDG1 is the surface brightness fluctuation (SBF) method. This was indeed attempted by \cite{Danieli2022} using imaging from HSC on the Subaru telescope. However, the data were of low S/N resulting in a very large uncertainty of $\pm$5 Mpc for a best distance of 21 Mpc. 

\begin{figure*}
    \centering
    \includegraphics[width=0.99\linewidth]{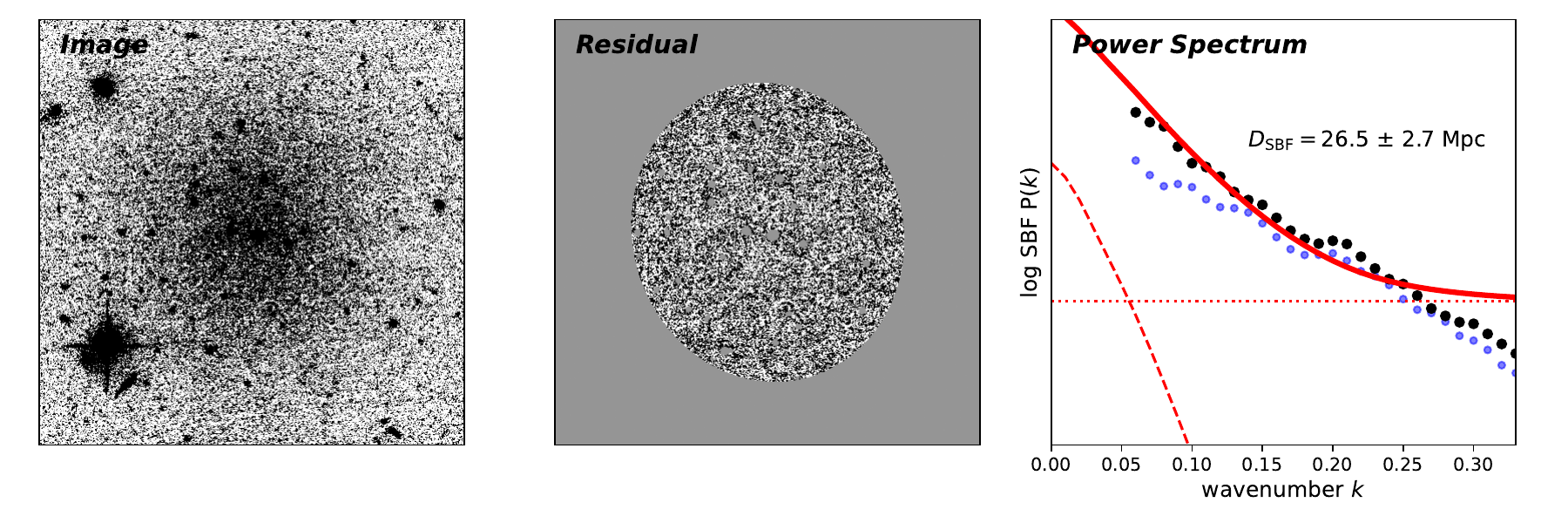}
    \caption{Surface brightness fluctuation distance for UDG1. {\it Left} the original HST/ACS F814 image. {\it Middle} the normalized residual image, with contaminants and the region beyond 1 R$_e$ masked. {\it Right} the azimuthally averaged power spectrum of the galaxy (black points), overlaid with the best-fitting model (red solid curve) with a combination of the PSF and white-noise components (shown individually with the red dashed and dotted lines, respectively). The contribution from unmasked background sources is shown by the blue points. The resulting distance to UDG1 of 26.5 $\pm$ 2.7 Mpc is indicated. 
}
    \label{fig:SBF}
\end{figure*}

Following the method of \cite{Tang2025b} we
use the HST/ACS F814W image, since it has a superior flat correction to the HST/WFC3 data, to derive the SBF distance to UDG1.
First, the 2D light distribution of this galaxy is modeled using GALFIT 
\citep{2002AJ....124..266P}, with a combination of a single Sersic profile and a sky background plane. We normalize the residual image by dividing by the square root of the best-fit Sersic model. The bright contaminating sources are masked, and the analysis is restricted to the region within the 1 R$_e$ ellipse to achieve higher S/N. We derive the azimuthally-averaged power spectrum from the normalized residual image and fit it as the sum of a PSF component and a constant white noise term. The SBF signal is equal to the amplitude of the PSF component. The same analysis is repeated in a blank field near the galaxy to estimate the SBF signal contributed by the background  which we subtract from the signal of UDG1. The SBF signal uncertainty is derived from the standard deviation of 1000 Monte Carlo realizations, generated by injecting Gaussian noise based on the error map. In Fig.~\ref{fig:SBF} we show the fit to the SBF signal.

To convert the SBF signal into distance, we adopt the empirical calibration for low-luminosity galaxies by 
\cite{2021ApJ...923..152K}. We use the stellar population model FSPS of \cite{2009ApJ...699..486C} to transform our HST values into the Subaru 
HSC 
system, which is required by the calibration of \cite{2021ApJ...923..152K}. The resulting SBF distance is  26.5 Mpc. The uncertainty is $\pm$2.7 Mpc. This uncertainty is primarily driven by the low surface brightness of UDG1 in the F814W image (i.e. the signal uncertainty) but the calibration uncertainty also contributes. 
We note that \cite{2001MNRAS.327.1004B} derive a Malmquist-corrected SBF distance to the central group galaxy itself (NGC 5846) of 25.6$\pm$2.4 Mpc. 
Our SBF distance measure is the same as the group distance of \cite{2017ApJ...843...16K} as assumed by \cite{Danieli2022} and over 2$\sigma$ different than the 20.0 Mpc claimed by \cite{2025arXiv251209990G}.

\section{An alternative approach to determine the number of GCs associated with UDG1}

As mentioned earlier, the HST/WFC3 observations of UDG1 were deep enough to essentially detect all of its expected GCs. Despite this, the two HST/WFC3 studies reached quite different total GC counts (i.e., 54$\pm$9 vs 33$\pm$3).

If the distance is known to a galaxy, then a common method to estimate the total number of GCs is to fit a Gaussian GCLF to the bright half of the system. After correcting for any contaminants, the total GC count is simply double the number of GCs brighter than turnover. 
In such cases the distance (and inferred apparent magnitude of the GCLF turnover) is an important factor in the accuracy of the total count which is  rarely reflected in the quoted count uncertainty.  
Using our measured SBF distance we can adopt the standard approach of 
using only the brighter half of the GCLF, which has the advantage of working with the higher S/N objects while avoiding the fainter ones for which contamination is an increasing concern. 



For our SBF distance of 26.5 $\pm$ 2.7 Mpc,  the turnover magnitude corresponding to M$_V$ = --7.5 
\citep{2003LNP...635..281R}
 is V = 24.62 (we discuss the use of a different turnover magnitude below). 
To convert the V band (Vega) into F606 (AB) we use the formula:
F606 = V -- 0.15(V--I), where V--I = 0.9 from \cite{2018AJ....156..296H}. This gives a turnover of F606 = 24.48. 
For such magnitudes the contamination rate is expected to be very low. In addition, all GC candidates are all resolved by HST and the lists of both studies are dominated by confirmed GCs at these magnitudes. \cite{Danieli2022} estimated the contamination rate to be $\le$0.3 GCs for F606 $<$ 24.5. Here we assume the contamination rate is zero and that the GC counts brighter than the turnover can be simply doubled. 

Table C1 of \cite{2025arXiv251209990G} lists 36 GC candidates. When examining their GCLF they excluded the 3 GC candidates with star-like profiles, i.e. GC20, 33 and 35. 
Here we do the same (the remaining 33 candidates include two that lie beyond 2~R$_e$.) Counting the candidates brighter than, or equal, to the turnover magnitude of F606 = 24.48 gives 25 GCs. 
The number of 25 GCs is simply doubled to give a total of 50.
Repeating the exercise for the larger allowed distance (29.2 Mpc) and closer distance (23.8 Mpc), corresponds to turnovers of 24.69 (F606) and 24.25 (F606) respectively. Again from table C1, this gives a total of 54 GCs if UDG1 is located at 29.2 Mpc and 42 if located at 23.8 Mpc. 

Similarly, using the list of GC candidates supplied by S. Danieli (2025, priv. comm.) for objects within 2~R$_e$ we count 22 GCs brighter than the turnover magnitude of 24.48 (F606).
To get the final GC numbers we add 2 GCs to account for those that lie beyond 2~R$_e$ (same as done above). Thus there are a total of 24 GCs brighter than the turnover magnitude of 24.48 (F606), giving a total system of 48 GCs. For the larger distance the total is 60 and for the closer distance it is 40. 

\begin{figure}
    \centering
    \includegraphics[width=1.05\linewidth]{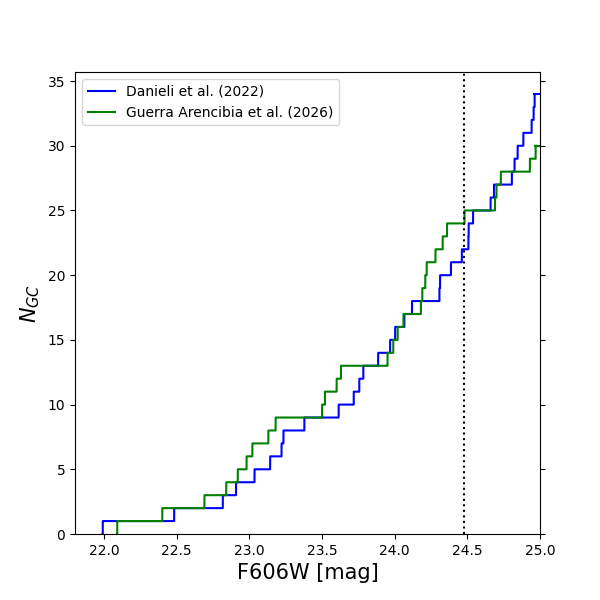}
    \caption{Cumulative distribution of GC number selected by 
    \protect\cite{Danieli2022} and 
    \protect\cite{2025arXiv251209990G} as a function of bright to faint F606 magnitude (blue and green, respectively). The plot shows the range of magnitudes for the resolved GCs, i.e. down to F606W = 25. The vertical line shows the GCLF turnover magnitude of F606 = 24.48 which corresponds to M$_V$ = --7.5 and a distance of 26.5 Mpc. 
    }
    \label{fig:cf}
\end{figure}

This process is illustrated in  Fig~\ref{fig:cf} in which we show the GCs selected by the two studies down to a magnitude of 25 (F606). 
Noting that two additional GCs should be added to the \cite{Danieli2022} distribution to account for those at large radii, the two distributions are quite similar and reach a similar total number of GCs. In particular, the two studies have similar numbers of GCs selected to the GCLF turnover of F606 = 24.48 (which corresponds to M$_V$ = --7.5 for a distance of 26.5 Mpc).

In this exercise we have used M$_V$ = --7.5 as the GCLF turnover magnitude. We acknowledge that other values have been used in the literature, e.g. 
M$_V$ = --7.6 \citep{2012Ap&SS.341..195R} and M$_V$ --7.3 \citep{2007ApJ...670.1074M}. 
It is not currently clear which is the appropriate turnover magnitude for UDGs like UDG1 which have dwarf-like stellar masses but giant-like halo masses \citep{2026arXiv260221875G}.
However, the other choices of turnover magnitude change the total GC number by $\le$ 6 GCs. 

In summary, using a distance of 26.5 $\pm$ 2.7 Mpc and doubling the number of GCs  brighter than, or equal to, the GCLF turnover magnitude of M$_V$ = --7.5 we find UDG1 hosts 
50$^{+4}_{-8}$ GCs taking those selected by \cite{2025arXiv251209990G} and 
48$^{+12}_{-8}$ GCs with those selected by \cite{Danieli2022}. Using this robust approach the two studies are entirely consistent with each other.
It also confirms a M$_{GC}$/M$_{\ast}$ ratio of $\sim$10\%. 
A total GC count of $\sim$50 implies that UDG1 has a massive halo of $>$ 10$^{11}$ M$_{\odot}$ \citep{Burkert2020}.  Having re-confirmed the richness of UDG1's GC system, the question of why UDGs can host much richer GC systems than their classical dwarf counterparts remains an open one  \citep{Forbes2025}.



As mentioned in the Introduction, the HST/ACS study of UDG1 by \cite{Marleau2021} derived a total number of GCs within 2~R$_e$ of 38 $\pm$ 7 (an additional 2-3 GCs may lie outside of 2~R$_e$). They assumed a closer distance of 20.3 Mpc and a GCLF with a turnover of M$_V$ = --7.3. Had they used a distance of 26.5 Mpc, the apparent magnitude would be 0.58 mag. fainter, which would lead to a significant increase in their total GC count. 

Finally, we note that \cite{2025arXiv251209990G} estimated the distance to UDG1 using the mean size of their GCs. They measured a mean GC effective radius of 0.0348 arcsec for a Gaussian model. This corresponds to a distance of 19.9 Mpc for a typical physical size of around 3.4 pc as found in the Milky Way. However, the expected mean size of GCs in UDGs is uncertain. The size of GCs is on average larger for metal-poor GCs and larger for those hosted in lower mass galaxies \citep{2005ApJ...634.1002J}. So a mean size for the GCs in UDG1 of $\sim$4.5 pc (for a distance of 26.5 Mpc) may be entirely reasonable.  

\section{Conclusions}

We revisit the GC system associated with the ultra diffuse galaxy NGC5846\_UDG1. In particular, focusing on the imaging studies of \cite{Danieli2022} and \cite{2025arXiv251209990G} who reported quite different GC counts for the galaxy, i.e.  54 $\pm$ 9 versus 33 $\pm$ 3 GCs respectively.  In both studies the GC candidates were identified and counted directly due to the depth of the HST imaging.
Thus, the total GC count was largely independent of the 
distance to UDG1 (which has been subject to debate). 

We find that the differences in total GC count are not due to magnitude or  radial incompleteness, GC system radial extent, the Galactic extinction applied, or the ellipticity of the GC candidates. We also find that the \cite{2025arXiv251209990G} total F606 magnitudes are systematically brighter than those of \cite{Danieli2022} ($\Delta$m = 0.09), and those of \cite{Muller2021} ($\Delta$m = 0.15). The main difference, but not the only one, is the more restricted colour criterion applied by \cite{2025arXiv251209990G}. We have examined 7 of the brightest  GCs excluded by \cite{2025arXiv251209990G} but included by \cite{Danieli2022}.
They are all resolved in the HST imaging (ruling out stars) with physical sizes and magnitudes consistent with GCs. They are almost perfectly round with GC-like appearances and projected within 2~R$_e$ of the galaxy centre. 
Contamination from background galaxies or interloper GCs is expected to be fewer than one object. We conclude that these objects are neither stars, background galaxies nor interloper GCs but bona fide GCs associated with UDG1 that should be selected. 

Here we present a new SBF-based distance using archive HST/ACS imaging in the F814W filter. Our distance of 26.5 $\pm$ 2.7 Mpc places UDG1 squarely within the NGC 5846 group. Knowing the distance also allows us to adopt the standard approach of deriving total GC counts by doubling the number of GCs brighter than the turnover of the Gaussian GC luminosity function. For UDG1 this has the advantage of using only resolved GC candidates, the bulk of which are confirmed by spectra.  
With this robust approach we derive totals of 
50$^{+4}_{-8}$ GCs from the \cite{2025arXiv251209990G} GC candidate list and 48$^{+12}_{-8}$ GCs from the \cite{Danieli2022} study. Thus the two studies are entirely consistent with each other when restricted to the brighter magnitudes. We conclude that UDG1 does indeed have a rich GC system of $\sim$50 GCs, which implies a massive dark matter halo of over 10$^{11}$ M$_{\odot}$.  

\section*{appendix}

\begin{figure*}
    \centering
    \includegraphics[width=0.99\linewidth]{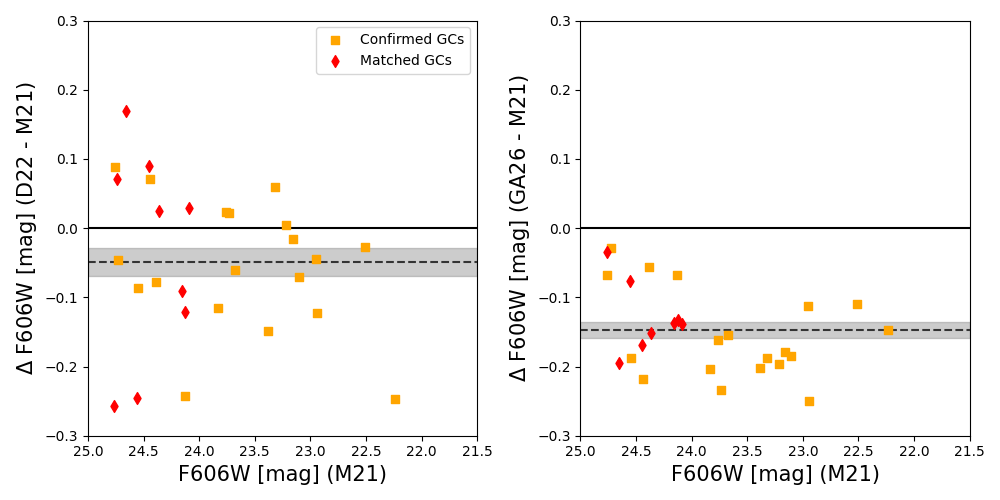}
    \caption{Difference in F606W magnitude for common GC candidates from 
    \protect\cite{2025arXiv251209990G} and 
    \protect\cite{Danieli2022} vs that of 
    \protect\cite{Muller2021}. Horizontal dashed lines show the mean offset and error on the mean. Different symbols show GCs that are in common (matched; diamonds) to both 
    \protect\cite{Danieli2022} and  \protect\cite{2025arXiv251209990G}, and those with spectra (confirmed; squares). 
    The \protect\cite{2025arXiv251209990G} magnitudes are systematically brighter by 0.15 mag., and  
    \protect\cite{Danieli2022} by 0.06 mag. brighter,  than the 
\protect\cite{Muller2021} photometry. A difference of $\sim$0.02 mag. is expected given that 
    \protect\cite{Muller2021} used ACS rather than WFC3. 
}
    \label{fig:mag}
\end{figure*}

In Fig.~\ref{fig:mag} we show  
total F606 magnitudes from \cite{Danieli2022} and \cite{2025arXiv251209990G} relative to the F606 magnitudes from table 3 of \cite{Muller2021} who used the ACS F606 filter.
We find 
\cite{2025arXiv251209990G} to be 0.15 $\pm$ 0.01  brighter and \cite{Danieli2022} to be 0.06 $\pm$ 0.02 mag. brighter than \cite{Muller2021}. Based on the differences between the ACS F606 filter used by 
 \cite{Muller2021} and the WFC3 F606 filter, both studies would be expected to be $\sim$0.02 mag. brighter. 
This suggests that either \cite{Danieli2022}, and \cite{Muller2021}, have underestimated the total GC fluxes or that 
\cite{2025arXiv251209990G} have overestimated them.  

\section*{Acknowledgements}

We the referee for their report which helped us to improve the paper. 
We thank the AGATE team for their input, including A. Levitskiy, J. Gannon, L. Haacke, D. Vaz, M. Monaci and H. Christie. We thank S. Guerra Arencibia for providing further details from their analysis. We thank S. Danieli for supplying a list of selected sources and for comments on the manuscript. 
We thank the ARC for financial support via DP250101673. DF dedicates this paper to Carl Grillmair, astronomer, colleague and friend. 

\section*{Data Availability}
This article is based on  work that is publicly available.



\bibliographystyle{mnras}
\bibliography{bibliography} 







\bsp	
\label{lastpage}
\end{document}